# Effects of Quantum Computing in Security

Abdulrahman AlRaimi
CSE Department
Qatar University
Doha, Qatar
aa1704278@qu.edu.qa

Sandrik Concepcion Das
CSE Department
Qatar University
Doha, Qatar
sd1706553@qu.edu.qa

Saad Mohammed Anis
CSE Department
Qatar University
Doha, Qatar
sa1703847@qu.edu.qa

Devrim UNAL
KINDI Center for Computing Research
Qatar University
Doha, Qatar
dunal@qu.edu.qa

*Abstract*— One of the fundamental theories of physics is that of quantum mechanics. Quantum mechanics tries to explain the inconsistencies in the behaviors of systems at the macro and micro scales. Quantum mechanics paved the way for quantum computing based on qubits. The existence of quantum computers up to 65 qubits is known. The advent of quantum computers weakens the security of many cryptographic algorithms. In this paper, we investigate quantum computing-based attacks and shed light on possible future developments.

## I. Introduction

Developed in the early 20th century, quantum mechanics explained the particle-wave duality of light and extended the phenomenon to electrons and atoms. Quantum physicists saw this as an exciting problem to leverage the quantum nature for advanced computation. Richard Feynman in 1981, Paul Benioff in 1979, and Yuri Manin in 1980 laid out their theories of quantum computing and how it should be helpful. In summary, Quantum computing is the technology that exploits specific properties of quantum mechanics to perform computation.

### A. Superposition

One of those properties is superposition. Quantum superposition is the property of quantum systems to exist in opposing states at the same time. This principle of quantum mechanics presents that systems only have a definite state when observed, before which they remain in an indeterminate state – in a combination of different states. The polarization of light demonstrates an example of this property. Light, commonly seen, is polarized in an arbitrary unspecified direction. The polarization of light is a combination of vertical and horizontal polarization:

$$|\psi\rangle = \alpha|\uparrow\rangle + \beta|\rightarrow\rangle$$

Where $|\psi\rangle$ is the polarization state, $|\uparrow\rangle$ and $|\rightarrow\rangle$ are the vertical and horizontal polarizations. and $\alpha$ and $\beta$ are amplitudes or the coefficient of each polarization. Polarizing filters filter light that is polarized parallel to the axis of the filter. Two filters placed one after the other, one vertical and the other horizontal, would produce no light after the second filter since $|\rightarrow\rangle$ is orthogonal to $|\uparrow\rangle$. But if the second filter is placed at a 45° angle, some light will get through to the other side. This is because the vertically polarized light is a superposition of two diagonal components:

$$|\uparrow\rangle = \frac{1}{\sqrt{2}}|\nearrow\rangle + \frac{1}{\sqrt{2}}|\nwarrow\rangle$$

According to the Born rule by Max Born in 1926, the modulus squares of the amplitude of a state gives the probability of the state being that value after measurement. The amplitude $\frac{1}{\sqrt{2}}$ in the example gives:

$$\left|\frac{1}{\sqrt{2}}\right|^2 = \frac{1}{2}$$

The value $\frac{1}{\sqrt{2}}$ is chosen to equate the sum of the modulus squared of the amplitudes to 1. Because the modulus squared of the amplitudes $\alpha$ and $\beta$ are probabilities, $\alpha$ and $\beta$ can both be complex numbers.

### B. Entanglement

Another property of quantum mechanics utilized in quantum computing is entanglement. In 1935, Albert Einstein, Boris Podolsky, and Nathan Rosen published a paper (now called EPR) which tried to rebuke quantum mechanics by showing the property of entanglement, which eventually became a significant property of quantum mechanics. Entanglement is a property demonstrated by two quantum systems that are inseparably correlated in their superpositions. The correlation of two entangled particles stores the information about the particles, instead of each particle storing its own information.

### C. Reversibility

Another fundamental principle of quantum computing is the exclusive use of reversible logical operations – operations that allow reverting to the input using the output, thus losing no information.

### D. Qubit

Classical computers run on classical bits - 0 or 1. Quantum computers use qubits. Like classical bits, a qubit can be a 0 or 1, but it can also be a value in a continuous range between 0 and 1 – a superposition of 0 and 1. The states $|0\rangle$ and $|1\rangle$, the computational basis of a two-level system, can be represented in the form of matrices:

$$|0\rangle = \begin{pmatrix} 1 \\ 0 \end{pmatrix}$$

$$|1\rangle = \begin{pmatrix} 0 \\ 1 \end{pmatrix}$$

### E. Quantum Operators

Operators used in quantum computers change the qubits in different ways to perform computation. These operators are unitary and reversible. Some operators are their own inverses – involutive, and some are not.[1]

## II. POSITIVE IMPACT OF QUANTUM COMPUTING

Although quantum mechanics will negatively affect cyber security, it introduces a world with exciting rules that could be exploited to our benefit. We call the era when we will use quantum mechanics everywhere the Post Quantum Era and its cryptography Quantum Cryptography. We could say that the largest positive impact of Quantum Mechanics is, putting our world of cyber security in danger. Because of this, we need to prepare for the Post Quantum Era by developing new protocols and encryption scheme that is resilient to Quantum and Classical computers. Quantum mechanics is the reason behind this research move, which will lead to more robust and better cyber security in the future. A good example is the B92 protocol, which will be explained in this paper. Before diving into B92, we will start by introducing light polarization.

### A. B92 Protocol

The first Quantum Key Distribution Protocol that saw the light is BB84 protocol by Charles Bennett and Gilles Brassard in 1984. B92 protocol has been developed based on BB84 protocol by Charles Bennett in 1992. The aim for B92 protocol is to make it easier and less complex than BB84 protocol.[3] To start explaining the B92 protocol, we need to know some basics. Simply, the detector of the photon has two bases, $+$ and $\times$. If the detector is adjusted to detect a photon on the $+$ base, it will detect either a vertically polarized light or a horizontally polarized light. If the detector is adjusted to detect a photon on the $\times$ base, it will detect a diagonally polarized light, either with $+45^o$ or $-45^o$. To start this protocol, Alice will use either a horizontally polarized photon ($\rightarrow$) which will represent the 0 value, or a diagonally polarized photon with $+45^o$ ($\nearrow$) which will represent the 1 value. In the other hand, Bob will use either $+$ or $\times$ base to detect the coming photon. But how will Bob determine the value of the coming photon?[17]

TABLE I. POLARIZED LIGHT AND DETECTORS

| Polarized Photon | Detector Base | Detected Photon | Value |
|---|---|---|---|
| $\rightarrow$ | $+$ | $\rightarrow$ | … |
| $\rightarrow$ | $\times$ | $\nearrow$ | … |
| $\rightarrow$ | $\times$ | $\searrow$ | 0 |
| $\nearrow$ | $+$ | $\uparrow$ | 1 |
| $\nearrow$ | $+$ | $\rightarrow$ | … |
| $\nearrow$ | $\times$ | $\nearrow$ | … |

Fig. 1. Combinations of polarized light and detector bases for the B92 protocol.

From figure 1, we understand the following:

- If a photon is detected using the $+$ base, there are 2 options:
  - If $\rightarrow$ is detected, then either $\rightarrow$ or $\nearrow$ has been sent. So, we cannot determine the value.
  - If $\uparrow$ is detected, then 100% $\nearrow$ has been sent. So, the value is 1.
- If a photon is detected using the $\times$ base, there are 2 options:
  - If $\nearrow$ is detected, then either $\rightarrow$ or $\nearrow$ has been sent. So, we cannot determine the value.
  - If $\searrow$ is detected, then 100% $\rightarrow$ has been sent. So, the value is 0.

So, a transmission of a secret key will be like this:

- For each photon, Alice Polarize it horizontally or diagonally with $+45^o$.
- Alice sends all photons to Bob.
- Bob detects each photon with $+$ or $\times$ base.
  - If the detected photon is $\uparrow$, then the value is 1.
  - If the detected photon is $\searrow$, then the value is 0.
  - Otherwise, no value is assigned.
- Bob tells Alice the result of his measurements.
- Alice and Bob retain the corrected measured photons and discards the rest.

### B. Improved B92 Protocol with Pulse Position Modulation

It has been proved that the effective transmission for B92 Protocol is 25%, meaning that 75% of the measurement is discarded, which means that the key rate is very low. So, how could we improve the key rate? This could be achieved by using pulse position modulation.[3] We will not discuss this technology here, but we will discuss its effect. Using PPM, we could transform each bit to represent n bits. The idea is very simple. We will choose first how many bits we want each optical pulse to represent. Say we chose 2. Then, it was proved that we need at least 2*n time slots to represent n bits. In our case, we need 4 time slots. The idea is to have a frame and divide it into 2n time slots, therefore, 4 time slots in our case. Then encode the optical pulse in one of these time slots. Each time slot represents a value that could be represented using two bits. Imagine you are a receiver, for each frame, you will have 4 possible values. If the optical pulse is encoded in:

- 1st time slot: the value is: 0 → 00
- 2nd time slot: the value is: 1 → 01
- 3rd time slot: the value is: 2 → 10
- 4th time slot: the value is: 3 → 11

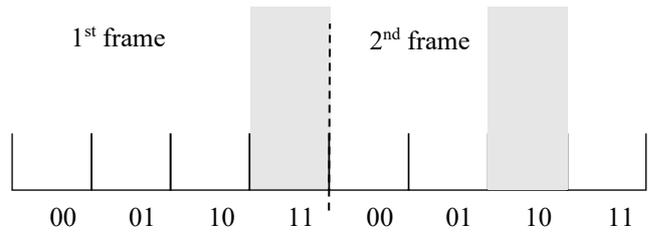

Fig. 2. A modulated frame using 4-PPM technology.

A transmission of multiple photons in the B92 Protocol with the 4-PPM will be like this:

- For each optical pulse, Alice prepares a frame with 4 time slots.
- Alice polarizes each optical pulse horizontally or diagonally with $+45^o$.
- Each polarized optical pulse will be encoded on a random time slot on its associated frame.
- Alice sends all frames to Bob.
- Bob detects each pulse with $+$ or $\times$ base.
  - If the detected pulse is $\uparrow$ or $\searrow$ then get the value using 4-PPM rules.

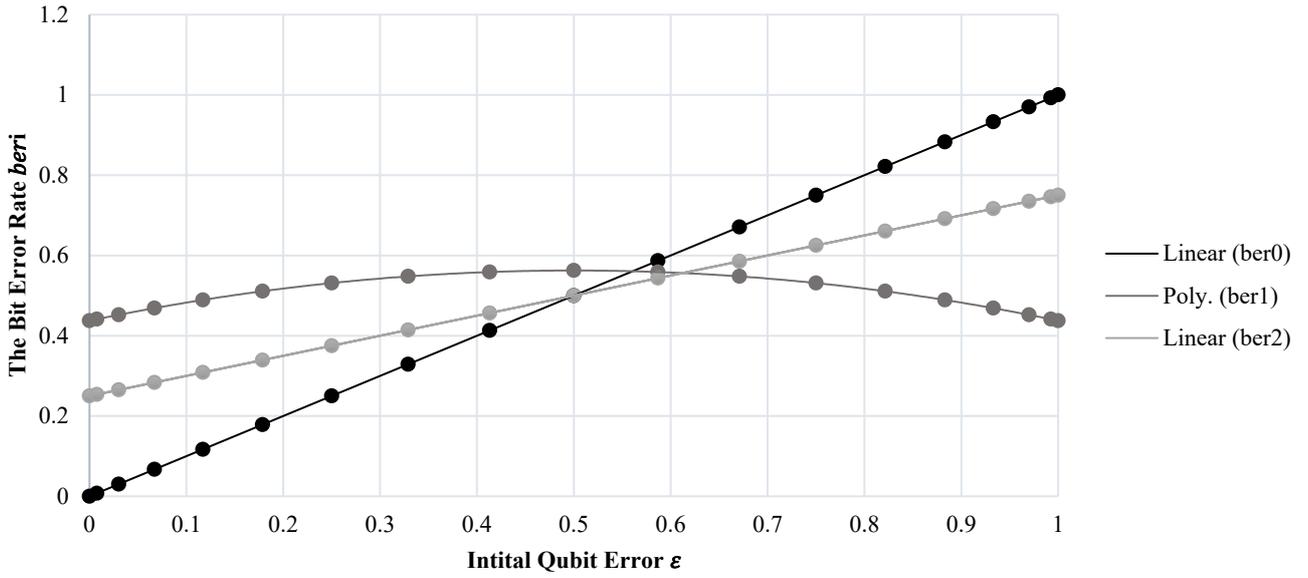

Fig. 3. Collective Rotaion Noise Channel

- Otherwise, no value is assigned.
- Bob tells Alice the result of his measurements.
- Alice and Bob retain the corrected measured photons and discards the rest.

*C. Security of the B92 Protocol*

To transfer a light particle between Alice and Bob, we need a channel. A lossy channel is an ideal channel where there is no noise. Noise in a channel is something that causes a bit error. An Example of bit error is when Alice sends 1 while Bob detects 0 or vice versa. It was proven that B92 protocol is unconditionally secure, and any attempts of eavesdropping are detected in a lossy channel.[5] However, we know that in real life, nothing is perfect. So, we need to analyze the security of B92 in a noisy channel.

*D. Collective-Rotation Noise Channel*

One type of noisy channel is the Collective-Rotation noise channel. A particle transmitting through this channel will deflect with angle $\theta$. One thing to note is no direct way to distinguish between eavesdropping error and noise error. So, in order to detect eavesdropping, we need to determine the initial qubit error rate $\varepsilon$. The initial qubit error is the maximum noise rate that could happen to a particle in the absence of eavesdropping. If noise rate has been computed for a certain communication and it was higher than $\varepsilon$, then an eavesdropping act has occurred 100%.[4] Before diving into the result of this analysis, we introduce these variables:

- $ber_0$ is the initial qubit error $\varepsilon$.
- $ber_1$ is the bit error rate when the eavesdropper detects a photon successfully, either ↘ $or$ ↑, and resend it directly to Bob.
- $ber_2$ is the bit error rate when the eavesdropper detects a photon successfully, either ↘ $or$ ↑, and send either ↗ $or$ → to Bob.

Since detailed proofs are behind the scope of this paper, we will introduce the results directly. After the analysis, we observe the following:

- $ber_0 = sin^2(\theta) = \varepsilon$
- $ber_1 = \frac{7+8\varepsilon(1-\varepsilon)}{16}$
- $ber_2 = \frac{1}{4} + \frac{1}{2}\varepsilon$

From figure 3, there are two cases:

- If $0.5 \leq \varepsilon \leq 0.56$, then $ber_1 \geq ber_0 \geq ber_2$
- If $\varepsilon \leq 0.5$, then $ber_1 \geq ber_2 \geq ber_0$

From point 1, we understand that we could detect the eavesdroping act only if Eve resends what she detected directly. If Eve chooses always to send either ↘ $or$ ↑, no way to detect it. Therefore, a channel with an initial qubit error greater than 0.5 must be avoided because the noise rate is very high, and we cannot detect Eve. However, we understand from point 2 that a channel with an initial qubit error smaller than 0.5 is secure because any eavesdropping act will increase the noise rate higher than the initial qubit error, so Eve will always be detected. Assume an eavesdropping act existed on a channel with an $\varepsilon \leq 0.5$; the most amount of information Eve can gain will be 50%, and she will not be able to know what she gets. Even if she knows what she gets, she only got 50% of the key. Hence, the B92 protocol is secure when $\varepsilon \leq 0.5$.

III. NEGATIVE IMPACT OF QUANTUM COMPUTING

While quantum computing shows high potential towards the advancements in computational processing and simulation, the same power it possesses also renders dangerous shortcomings.

*A. Issues with quantum particles*

One of the issues that arise from the properties of quantum bits, also known as "qubits" is called quantum decoherence. Decoherence is the loss of the ability of qubits to remain in quantum superposition and/or entanglement. Qubits are very sensitive to their environment and any slight disturbances could result in decoherent qubits, which essential lose their ability to be in a bi-state superposition. Hence, quantum computers are specifically built in void conditions such as sub-zero temperatures and vacuumed spaces.[6]

### B. Vulnerability of existing cryptographic techniques

*1) Asymmetric encryption:* Asymmetric public key encryption methods such as the Rivest, Shamir, and Adleman (RSA) and the Diffie-Hellman (DH), which rely on the principles of discrete logarithms and large integer factorization being hard to decipher, can be easily dissolved on quantum computers. A quantum technique recognized as Shor's algorithm can quickly render these algorithms vulnerable through the process. Shor's algorithm works around the principle that qubits can hold a superposition of all possible values (in this case, prime factors), thus exposing the two private numbers following specific computational steps.

*2) Symmetric encryption:* With respect to symmetric encryption, which uses the same private key and cipher to encrypt and decrypt information (bidirectional), an algorithm developed by Lov Grover can be implemented on a quantum computer to find a specific target element within the order of $O(\sqrt{N})$ operation, compared to a conventional classical computer which could complete the search in O(N) steps. For example, working with 128-bit key length would be around the security level of a 64-bit key length cipher. However, this vulnerability can be slightly mitigated by increasing the key sizes by two times minimum.

*3) Hash functions:* Generally, a hash function convert a large random set of inputs into a fixed set output and is known to operate in a unidirectional manner. Since the produced output is of a fixed size, Grover's algorithm can detect a collision in $O(\sqrt{N})$ steps, similar to popular symmetric ciphers. Nonetheless, hashes such as SHA-2 and SHA-3 stand resistant to quantum techniques.[2]

## IV. GROVER'S ALGORITHM

Symmetric encryption is an encryption scheme where Alice encrypts a message with a key, and Bob decrypts this message with the same key. One way to find the key is the exhaustive search which could be achieved by trying all possible keys. For example, a message $p$ has been encrypted with a key $k$ of 16-bits which produced a ciphertext $c$. In exhaustive search, we need to try $2^{16} - 1$ keys in the worst case. So, the time complexity for exhaustive search is $O(2^n)$ where $n$ is the key size. In today's world, Advanced Encryption System (AES) is integrated into most applications as the main way to encrypt messages. In AES, a message could be encrypted with either 128-bit, 192-bit, and 256-bit key sizes. Without a doubt, there is no classical computer in the world that could find a key for a message that has been encrypted using AES-128 in a reasonable amount of time. The best-known attack on full AES is the key recovery attack which needs to try $2^{126}$ keys for AES-128, which is still infeasible. However, this is only the case with a classical computer. In Quantum world, Grover Algorithm can find the key with just $O(\sqrt{2^n})$. So, instead of trying $2^{128}$ possible keys, we will try $2^{64}$ possible keys. To understand how huge this improvement is, think about the ratio between these two numbers:

$$2^{64} : 2^{128}$$
$$1 : \sim 18 \times 10^{18}$$

Therefore, for each 1 key that will be tested in Quantum Computer, $18 \times 10^{18}$ keys will be tested in classical computer. Which means that Grover Algorithm is an enormous success and a dangerous threat to the classical symmetric encryption schemes in general.

### A. The Heartcore of Grover's Algorithm.

Grover Algorithm is a general quantum search algorithm for an unstructured database. The core for Grover Algorithm is the Oracle Function. To understand what Oracle function does, imagine we have a list from 1 to 4. In this example, we need to find the number 3. The Oracle function is a function that produces 1 when we put a solution and 0 otherwise. So,

$$f(x) = 0, x \neq 3 \text{ and } f(3) = 1$$

Generally, the oracle function is defined as follows:

$$f(x) = 1, \text{when } x = a \text{ } (a \text{ } solution)$$
$$f(x) = 0, otherwise \text{ } (no \text{ } solution)$$

For every certain problem, we need to create an oracle function that follows the criteria above and integrate it to a Grover Algorithm.[15] But what are the steps of Grover Algorithm? Imagine we have list from 1 to 4, and we need to find number 3.

### B. Grover Algorithm Steps

Firstly, how may qubits we need? Notice that we can represent each number in our list in terms of 2 bits, 00, 01, 10, 11. So, we need in total 2 Qubits.[14]

*1) Initilazation Step:* We need first to initialize the 2 Qubit to zero then we put each Qubit in the superposition state. By doing this, we are giving each possible solution the same probability to be chosen. For example, the probability of choosing 00 is 25% just as the probability of choosing 01, 10, or 11.

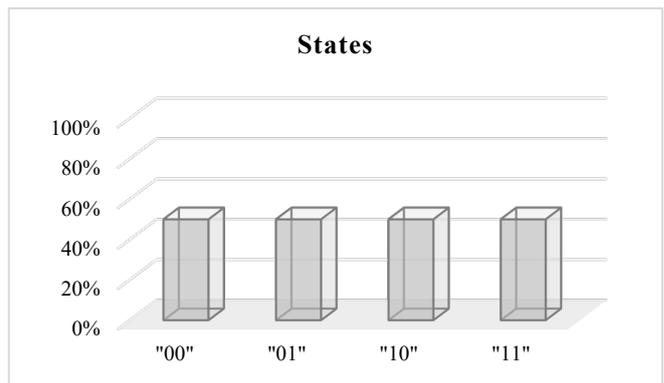

Fig. 4 Initilazation Step

*2) Oracle Function Step:* All the Qubits, which are in superposition state, will enter the oracle function. The oracle function will simply "mark" the most possible solution. In our case the solution is 11.

*3) Diffusion Operator Step:* Since we have a state that could be a solution, now we need to give more probability of

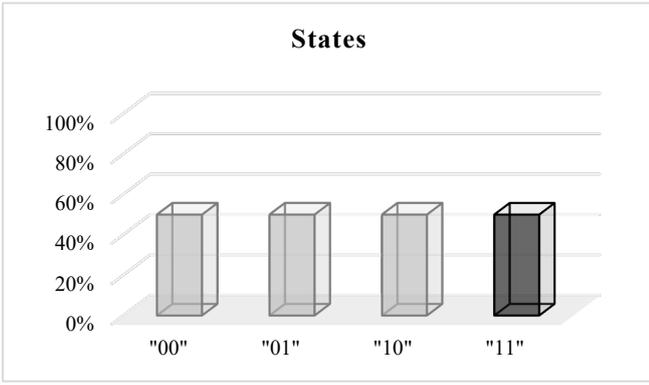

Fig. 5. Oracle Function Step

success. To do this, we will use something called Grover diffusion operator, which does this exactly.

 4) *Repetition:* We will simple repeat steps 2 and step 3 $\sqrt{2^n}$ times. In our case, we will repeat the steps $\sqrt{2^2} = 2$.

 5) *Measurement:* This is the final step. In this step, we

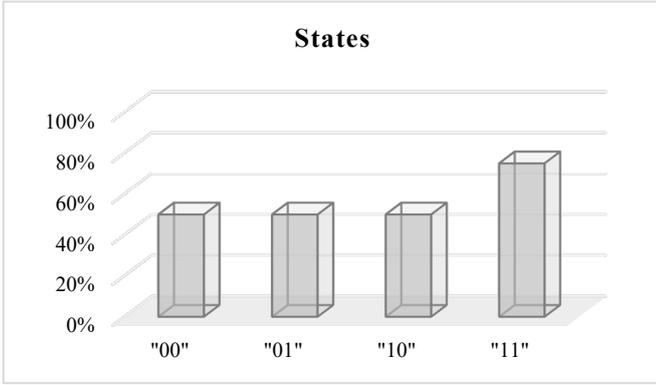

Fig.6 Diffusion Operator Step

will meaaure the result. In our case, the solution will be 11 with a probability that is near 100%.

*C. Grover Algorithm on Simplified AES*

It has been shown that Grover Algorithm could be used to find a key for a plaintext that have been encrypted using a simplified version of AES called SAES. In SAES, the key size is 16 bits. One might say, it is only 16 bits, where is the risk? The answer is that it could theoretically enlarge to any number of bits with only one limitation: Hardware. The main reason why researchers underestimate the value of the Grover algorithm is that you need either a Universal Quantum Computer or AES Quantum Circuit integrated into a Quantum computer for that specific purpose. Universal Quantum Computers do not exist. However, one AES Quantum Circuit has been built. The purpose of an AES Quantum Circuit is that an adversary can integrate it later as an oracle function. Considering that we need to build a SAES with the lowest number of Qubits, we needed around 64 Qubits. Results have been shown that an adversary will find the key for a certain text that has been encrypted with AES after completing $\frac{\pi}{4}\sqrt{2^n}$ iterations only. So, theoretically, AES is in danger. However, in real life, we are still in the early stages of Quantum Computing. Meaning, we might see the benefit of the Grover Algorithm, or any similar algorithm after a decade or so. However, we need to prepare for its effects from now.[7]

## V. SHOR'S ALGORITHM

The implications of using quantum computing, particularly in cryptography, are speeding up the process not through lower resource demand but rather by reducing the number of steps during computation. As mentioned in earlier sections, the heart of asymmetric encryption methods, such as the RSA, is based on the significant resource demand to factorize large integers.

*A. Shor's Algorithm Steps:*

To better understand the process, the classical procedure of integer factorization using Shor's algorithm is explained in the following subsection:[8] Let us assume that N, the number we are trying to factorize, has only two factors, p and q.

 1) *Step 1*: Start with a guess value, a, such that $a < N$, and a and N are co-prime, i.e., they do not share any common factors with each other. If $gcd(a,N) \neq 1$, therefore, a is a factor of N, and we can perform normal division, $N \div a$, to find the other factor, hence solving the factorization.

 2) *Step 2*: Compute the period of the function,
$$f_{(a,N)}(x) = a^x \bmod N.$$
Call this period r. If r is odd, redo step 1.

 3) *Step 3*: Since the period r is assumed to be even, check:
$$a^{\frac{r}{2}} + 1 \neq 0 \bmod N.$$

 4) *Step 4:* We know that $a^r = 1 \bmod N$. Therefore, $a^r - 1 = 0 \bmod N$. Also, For some given multiple k of N, $a^r - 1 = k \times N$. Also, since r is even, therefore:
$$\left(a^{\frac{r}{2}} - 1\right)\left(a^{\frac{r}{2}} + 1\right) = k \times N = k \cdot (p \cdot q).$$
We can then assume that:
$$p = gcd\left(a^{\frac{r}{2}} - 1, N\right) \text{ and } q = gcd\left(a^{\frac{r}{2}} + 1, N\right).$$

*B. Quantum Implementation*

The quantum implementation of Shor's algorithm mainly requires two co-entangled quantum registers, R1 and R2. Register R1 will roughly contain an adequate number of qubits needed to represent the integers from 0 until N-1 in a superposition or until a noticeable pattern can be observed to find the period. The function $f_{(a,N)}(x) = a^x \bmod N$ will be evaluated for all values of x till N, and the results will be stored in the register R2.[13] The rest of the implementation is fundamentally identical to the classical procedure since the main point of the quantum implementation is to calculate the period of the stated modulo function in a shorter number of operations. The implementation also involves techniques, such as applying the Hadamard gate, black-box operation, and quantum Fourier transformation onto register R1.[9]

## VI. FUTURE RESEARCH TRACK

With regards to the coming decades, compact quantum computers seem to be out of reach due to hardware and environmental limitations. Since qubits have short coherent times, therefore the short effective usage time to work with them drastically hinders processing complex computations. Another field that is open to further research is error correction or fault-tolerant quantum systems. This sector holds significant priority for development since qubits are hypersensitive to their environments and measurement of quantum readings are done through unorthodox protocols. Moreover, simply increasing the number of qubits does not contribute towards better results without lowering their error probabilities.

Despite all these challenges, enterprises like IBM and Google have ambitious plans to upscale quantum computers. As of late September 2020, IBM's current largest quantum computer contains 65 qubits and looking towards developing a 1000 qubit system by 2023. Such a landmark would be able to mitigate the fundamental issues in the current rudimentary state of quantum technology and would pave the path towards quantum supremacy.[10] Researchers have also theorized fault-rectifying conventions that involve multiple Qubit encoding from a single qubit.[11]

Another department that could use further research is cryogenic semiconductor technology. Since some quantum architectures rely on Quchips (short for "quantum chips") consisting of qubits, control circuits, and measuring devices, it is most likely that they are in a cryogenic setting altogether. This brings the need for cryogenic semiconductor technologies that fulfill the required functionalities that can withstand low temperatures without inducing internal system noise and producing erroneous qubits.[12]

Most of the above discussed security mechanisms based on quantum computing are resource and energy intensive, therefore they cannot be applied onto IoT systems due to resource and computational limitations [17]. IoT systems have high mobility which require specific computational models for modeling security aspects [18]. Especially, security policy models need to be designed and implemented according to the requirements of IoT devices, to provide secure devices without reducing its computational performance [19]. There is not much current work in the area of security modeling and security policy verification for quantum computing on mobile systems.

## VII. CONCLUSION

Quantum computing holds the promise to revolutionize the state of modern computing and encryption. However, quantum technology still requires decades of evolution and concentration to be integrated into daily life. The theory is developing steadily as multiple algorithms have been designed to solve particular problems more efficiently than classical computers. These algorithms have built the base on which further research could expand and meet future demands. Hardware developments, however, did not catch up with the speed of the theoretical work. The existence of a universal quantum computer is still an exceptionally far-off target. One of the significant obstacles in quantum computing is noise in quantum channels – a practical issue that ideal quantum computers do not possess. The developed algorithms assumed to be working on optimal machines strike an imbalance with the noisy practical examples. The power of quantum computing can help the world to improve computation astronomically, but it also holds a considerable risk of damage to the modern world. This technology, as researchers, should be handled, discovered, and exploited carefully to reap its eventual reward. With the technology steadily blooming, it is a great time to delve in as we expand our horizons into unparalleled computing power.